\documentclass[aps,prl,twocolumn,groupedaddress,showpacs]{revtex4}
\usepackage{graphicx}

\begin{document}

\title{The lower bound of barrier-energy in spin glasses: a calculation of the exponent on hierarchical lattice}

\author{A. Bhattacharyay}
\affiliation{Dipartimento di Fisica 'G. Galilei', Universit{\`a} di Padova,\\ Via Marzolo 8, 35131 Padova, Italy.\\}

\date{\today}

\begin{abstract}
We argue that the lower bound to the barrier energy to flip an up/down spin domain embedded in a down/up spin environment for Ising spin glass is independent of the size of the system. The argument shows the existence of at least one dynamical way through which it is possible to bypass local maxima in the phase space. For an arbitrary case where one flips any cluster of spin of size $l$, we have numerically calculated a lower bound to the exponent $\psi$ characterizing the barrier one has to overcome. In this case $\psi$ corresponding to the lower bound calculated on hierarchical lattice comes out to be equal to $\theta $ the exponent characterizing the domain wall energy in ground state. 
\end{abstract}


\maketitle

The slow dynamics of spin glasses [{\bf SG}] after a quench from a temperature greater then $T_c$ to one below $T_c$ ($T_c$ - spin glass transition temperature) is to a good extent explained under the assumption of droplet theory \cite{fish1,fish2} of coherent excitations of spin domains. The aging of a spin glass system (below $T_c$) is understood as a slow increase in coherence length $L_T(t)$ which occurs by accumulating smaller domains into bigger ones and in that way reducing the total domain wall length in the system. This, very slow non-equilibrium behavior which leads the system to-wards equilibrium is in excellent agreement with the scaling law \cite{fish1,jons}
\begin{equation}
L_T(t) \sim \left ( \frac{T{ln[t]_{\tau_0(T)}}}{\bigtriangleup(T)}\right )^{1/\psi},
\end{equation}
where $t$ is time with $\tau_0(T)$ being the time scale, $T$ is the temperature of the system whose energy has been calculated in the unit of Boltzmann constant $k_B$ and $\psi$ is the exponent associated with the free energy barrier one has to overcome in order to go from one minimum to another. It is taken that the barrier energy scales as $l^{\psi}$ where $l$ is the system size. The barrier exponent is considered to satisfy the relation $\theta\le \psi \le d-1$. The upper bound comes because of the fact that any compact cluster of length scale $l$ and energy less than $l^{d-1}$ can be created with a maximum domain wall size $l^{d-1}$ by simply moving the domain wall through the cluster. and the lower bound corresponds to the fact that one cannot flip a domain of size $l$ without encountering one of size at least $l/2$ whose free energy scales as $(l/2)^\theta$ \cite{fish2}. The exponent $\theta$ being much smaller than $(d-1)$ characterizes the free energy of a spin glass domain of linear size $l$. A lot of experimental end a few numerical work \cite{jons,dup,bou,mat,schi,bert,berth,gow} have shown the value of $\psi$ to fall within the above mentioned range. The fact that some experimentally evaluated value of $\psi$ in two and three dimensions is close to its upper limit is recently raising some question about the validity of the lower limit of $\psi$ set by Fisher and Huse \cite{fish2}. In a recent paper it has been argued in favor of a lower limit as $\psi=d-1$ \cite{bar} rather than $\psi=\theta$. An efficient numerical scheme has been proposed in Ref.-\cite{bar} following some underestimations relevant to the lower bound to barrier energy. In the present work we will mainly follow this numerical procedure (applicable for very high dimensions and length scales) but to definitely get different results than that in \cite{bar}.

\par
Getting an idea about the lower bound of $\psi$ in high length scales is a notoriously complicated affair in view of the fact that one ideally requires local optimizations of all possible spin flipping moves which is an NP-complete \cite{midd} process. Let us first present a dynamical way of sequential small excitation and subsequent stabilization which is able to slowly shrink the domain wall covering an entirely up/down spin domain embedded in a down/up spin environment. Fro simplicity, consider a rectangle separated by a straight line down the middle into two parts. In one part there are spins all in up-wards direction and in the other part there are spins all in the down wards direction. The middle line will be pushed to-wards one edge in such a way so that in the end the whole rectangular region is occupied by down spins only. In reality the domain wall is definitely not a straight line but much thicker than but our result can easily be generalized to that. Let us think about any sequence of single spin flip sequence that pushes the middle st line near its one end to-wards the up-spin domain living behind a very small extension of the down spin domain. Our changed domain wall i.e. the small pushed in part of the straight line, now has many spins frustrated on it immediately following the change in shape and we stabilize them to attain the ground state in the changed configuration of the spin domains. Since, in the above mentioned process we have only excited a very small portion of the domain wall and moreover, after the stabilization in the new state the domain wall length has changed only by a very small amount, the barrier energy encountered is very small. Now, we repeat the process of pushing the straight line (domain wall) in the same way just in the adjacent part of the previously pushed one and do the whole process. In the process, still we keep having the same stabilized length of the domain wall as in the previous state and had only excited a small part of it. So in this process we can make the down spin state encroach the up spin region totally but never going at a very high energy state compared to the initial one.
\par
Now, to generalize the above mentioned dynamics, consider an up/down spin domain completely embedded in a down/up spin one having a domain wall in the middle which can have a complex shape. Flipping the middle up/down spin domain can be achieved by slowly shrinking the domain wall in the above mentioned way to eventually make it vanish. In such a process since we are shrinking the boundary wall of a simply connected domain, we might at worst need to increase the domain wall by a very small amount. Moreover, since the excitations at each step of movement are kept very small the barrier state energy should never be very large compared to the initial state. If the up/down spin domain is multiply connected - meaning that there are droplets of oppositely oriented spins in the middle - the trick will be to flip all the smaller droplets inside following the same procedure to make the inner domain walls vanish first and then make the bigger domain wall go to zero as mentioned above. Thus, we can argue that there is at least one dynamical way corresponding to that mentioned above, where the barrier energy should not scale as the system size. For such a situation, there is definitely a way to bypass the local maximum in the phase space when one goes from one region to the other and the path corresponds to the lowest barrier path.         

\par 
But, definitely the above mentioned way is a too stricter one to follow considering thermal motion to produce it. In the following we would consider the way that involves a distribution of the excitations where larger domains can be flipped coherently at a time. Here one should think about optimizing on the flipping sequence of intermediate sized clusters to introduce or abolish a domain wall at a length scale $l$. In the rest of this paper we are going to show that the lower bound to $\psi$ is equal to $\theta$ for an Ising spin glass on a Migdal-Kadanof $[{\bf MK}]$ Hierarchical lattice. In what follows we apply renormalization group transformations to directly calculate $\psi$ at its lower limit at various length scales in many dimensions and will show the agreement in them at lowest and highest length scales (in our numerics) in all dimensions starting from $d=2$. Keeping in mind that it is an impossible task to probe all the relevant single spin flip sequences to make sure that one of it passes through the true barrier state one has to apply the optimized scheme of sequence of cluster flips with a cluster length-scale distribution appropriate to the given length scale of the system. Let us think about optimizing on a sequence of flipping cluster of spins before changing the boundary condition which will either introduce or remove a domain wall inside the system of size $l$. In such a process we would typically encounter the barrier state just before or after we change the boundary condition. When we flip a single cluster of spins which was previously in ground state in the process of flipping a sequence of them before changing the boundary conditions we suddenly frustrate a lot of spins on the previous domain wall of that cluster. In principle, one could find a single spin flip sequence for the spins on the domain wall before flipping the cluster as a whole which would at best lead one to a barrier state corresponding to the flip of that cluster with half as high energy. This is so because in this case also one would encounter the barrier state just before or after flipping the cluster followed by adjustment of spins on the domain wall. The error in this estimate will be of the order of energy associated with a single spin flip. Thus, the scope of adjustment of spins before flipping an intermediate sized cluster in a sequence of cluster flips to get to the barrier state associated with change in boundary conditions of a system of size $l$ could result in a barrier state half as high and will not essentially alter the scaling law. Since the conditions remain unchanged for spins lying inside the cluster (not on or adjacent to domain wall) undergoing flip we do not bother about them. Here, we actually consider that within cluster adjustments of spins after its flip as a whole are much low energy affairs to change our result.  
\par
Our numerical calculations are at $T=0$ on Ising spin glass. An Ising spin glass has the Hamiltonian
\begin{equation}
H = -\sum_{\langle ij\rangle} J_{ij}S_i S_j\, 
\end{equation}
and undergoes spin glass transition at nonzero $T_c$ in dimensions $d> 2$. In Eq.2 the sum is over all nearest neighbour pairs and the spins $S_i$ and $S_j$ can take up values $\pm 1$. There are good agreements of calculated value of $\theta$ on $MK$-hierarchical lattice and square and cubic lattice in two and three dimensions \cite{brey,marek,hart1,hart2} and a hierarchical lattice is widely in use for the numerical investigation of spin glasses. The $MK$-hierarchical lattice starts by forming in its first level a unit with $2^{d-1}$ parallel bridges connecting two end spins which set the boundary condition. Every bridge is a series connection of two bonds with a spin in the middle. In the present case at first level we select the bonds randomly from a pool of 10000 bonds generated from a Gaussian distribution of unit width and zero mean. In the next higher level of length scales one replaces the previous bonds with ones obtained from a renormalized bond distribution which is equivalent to replacing each bond in the previous level by the whole unit. Thus, in the $Ith$ level of iterations the renormalized lattice corresponds to a length scale $2^I$ where each spin in the middle is a domain of intermediate size. On such a lattice at $T=0$, the renormalization group transformation of effective bonds involves getting a contribution from each bridge as 
\begin{equation}
J_l = sign(J_1J_2)min[\vert J_1\vert,\vert J_2 \vert],
\end{equation}
where $J_1$ and $J_2$ are the bonds connected in series in the bridge. The renormalized bond is obtained by summing over contributions obtained from all parallel bridges.
\par
The basic theory we follow in optimizing the middle spin flip sequence before we flip the right hand corner spin to change the boundary condition is the same as in Ref.\cite{bar} apart from the fact that we are using renormalized bonds and are not concentrating to $L^{d-1}$ nearest neighbour spin to the right hand corner spin. Let us explain the steps of spin flip sequence following Ref.\cite{bar}. We would always go from the higher energy minimum to the lower one corresponding to the states with parallel and anti parallel end (corner) spins and will find the barrier height from the reference energy of the higher lying ground state. Since we are after finding the lower bound to the energy barrier the above mentioned move is justified. Now consider one of the $2^{d-1}$ bridges that connect the end spins. Let us take $e^1$ as the energy of the bridge measured from the reference of the previous ground state energy of it when the middle spin is flipped first before flipping the right hand corner one. Let $e^2$ be the energy of it when the corner spin if flipped first and the middle spin remains as it was and $e^3$ be the energy of the bridge when the middle spin is flipped first and then the corner spin is flipped. Taking the number of middle spins flipped before we flip the right hand corner spin to change the boundary conditions as $n$, the energy of the system just before and after flipping the corner spin be respectively 
 \begin{equation}
\sum_{l=1}^{n}{e_l}^1=E^1 
\end{equation}
and
 \begin{equation}
\sum_{k=2^{d-1}-n}^{2^{d-1}}{e_k}^2+\sum_{l=1}^{n}{e_l}^3=E^2 . 
\end{equation}
Since the energy barrier will be encountered just before or after flipping the end spin the barrier energy is given by
\begin{equation}
E=Max[E^1,E^2].
\end{equation}
As we are searching for the lowest barrier energy, we can make some underestimations. For a parameter $a$ satisfying $0<a<1$ we can write
\begin{eqnarray}\nonumber
E&=& Min_n[Max(E^1,E^2)]\\\nonumber
&=& Min_n[aE^1+(1-a)E^2]\\\nonumber
&=& Min_n\left[\sum_{l=1}^{n}{(a{e_l}^1+[1-a]{e_l}^3)}+\sum_{k=2^{d-1}-n}^{2^{d-1}}{e_k}^2\right ]\\
&=& \sum_{i=1}^{2^{d-1}}{Min[(ae^1+[1-a]e^3),(1-a)e^2]}.
\end{eqnarray}
\par
In the above expression setting $a=0.5$ means giving equal weight to $E^1$ and $E^2$ which will actually make the barrier energy negative in 2-dimensions \cite{bar}. This is because of the fact that we are always going from higher energy minimum to a lower energy minimum and it makes $e^2$ negative more often than not in 2-dimensions. In fig.1 we have plotted the log of barrier energy $E$ against log of system size ($l=2^I$) for $a=0.9$ . The plot shows graphs corresponding to dimensions from $d=2$ to $d=9$. We see all the graphs are straight lines. Here we selected $a=0.9$ because of the fact that small $a$ in the weights of $E^1$ and $E^2$ is not a good estimate for $d=2$ dimension. In Fig.2 we have plotted $\psi$ against all dimensions obtained at the lowest pair ($\psi_1$) and highest pair ($\psi_2$) of length scales in our numerical calculation. The two broken-line graphs corresponding to the two $\psi$ calculated at two extreme length scales almost fall on the continuous graph showing a plot of $\theta$ against same dimensions. 
\par
Now we would like to make a few comment regarding the difference of our result ($\psi = \theta $) at the lower bound and that obtained in \cite{bar} specially because we are using the simplified numerical scheme proposed in \cite{bar}. In \cite{bar} it has been tacitly taken that the barrier introduced or removed with the change in boundary condition passes through the nearest neighbour spins to the right hand corner one which is actually flipped to change the boundary condition. We argue that nobody knows where the domain wall will pass through corresponding to a change in the boundary condition and the domain wall width will actually be ignored if we concentrate on a sequence of spin flip as is done in \cite{bar}. Moreover, flip of a cluster of spins can be equivalently achieved by following sequences of single spin flip but there is no guarantee that one would get through a barrier state of lower energy unless a method of sequential excitation and stabilization is applied as has been explained in the introduction. All these things taken into consideration and specially when we do not know where the actual barrier will pass through or will be removed from when the boundary condition is changed, one must apply renormalization group approach to compare between the routes of going from one ground state to the other defined on appropriate length scales and the outcome is the same as that set by Fisher and Huse \cite{fish2}.

\acknowledgements{I acknowledge discussions with J.K. Bhattacharjee, Jayanth Banavar and Amos Maritan.}


\begin{thebibliography}{99}

\bibitem{fish1} D.S. Fisher and D.A. Huse, Phys. Rev. B
  \textbf{38}, 373 (1988).
\bibitem{fish2} D.S. Fisher and D.A. Huse, Phys. Rev. B
  \textbf{38}, 386 (1988).
\bibitem{jons} P.E. J\"onsson, H. Yoshino, P. Nordblad, H. Aruga
  Katori, and A. Ito, Phys. Rev. Lett. \textbf{88}, 257204 (2002).
\bibitem{dup} V. Dupuis, E. Vincent, J.-P. Bouchaud, J. Hammann, A. Ito, and H. Aruga Katori, Phys. Rev. B \textbf{64}, 174204 (2001).
\bibitem{bou} J.-P. Bouchaud, V. Dupuis, J. Hammann, and E. Vincent, Phys. Rev. B \textbf{65}, 024439, (2002).
\bibitem{mat} J. Mattsson, T. Jonsson, P. Nordblad, H. Aruga Katori, and A. Ito, Phys. Rev. Lett. \textbf{74}, 4305 (1995). 
\bibitem{schi} A.G. Schins, A.F.M. Arts, and H.W. de Wijn,
  Phys. Rev. Lett. \textbf{70}, 2340 (1993).
\bibitem{bert} F. Bert, V. Dupuis, E. Vincent, V. Hammann, and J.-P. Bouchaud, Phys. Rev. Lett. \textbf{92}, 167203, (2004).
\bibitem{berth} L. Berthier and J.-P. Bouchaud, Phys. Rev. B \textbf{66}, 054404, (2002).
\bibitem{gow} T.R. Gowron, M. Cieplak, and J.R. Banavar, J. Phys. A \textbf{24}, L127 (1991).
\bibitem{bar} B. Drossel and M.A. Moore, Phys. Rev. B \textbf{70}, 064412, (2004).
\bibitem{midd} A.A. Middleton, Phys. Rev. E \textbf{59}, 2571, (1999).

\bibitem{brey} A.J. Brey and M.A. Moore, J. Phys. C \textbf{17}, L463, (1984).
\bibitem{marek} M. Cieplak and J. Banavar, J. Phys. A \textbf{23}, 4385, (1990).
\bibitem{hart1} A.K. Hartmann, Phys. Rev. E \textbf{59}, 84, (1999).
\bibitem{hart2} A.K. Hartmann and M.A. Moore, Phys. Rev. Lett. \textbf{90}, 127201, (2003).
\end{thebibliography}
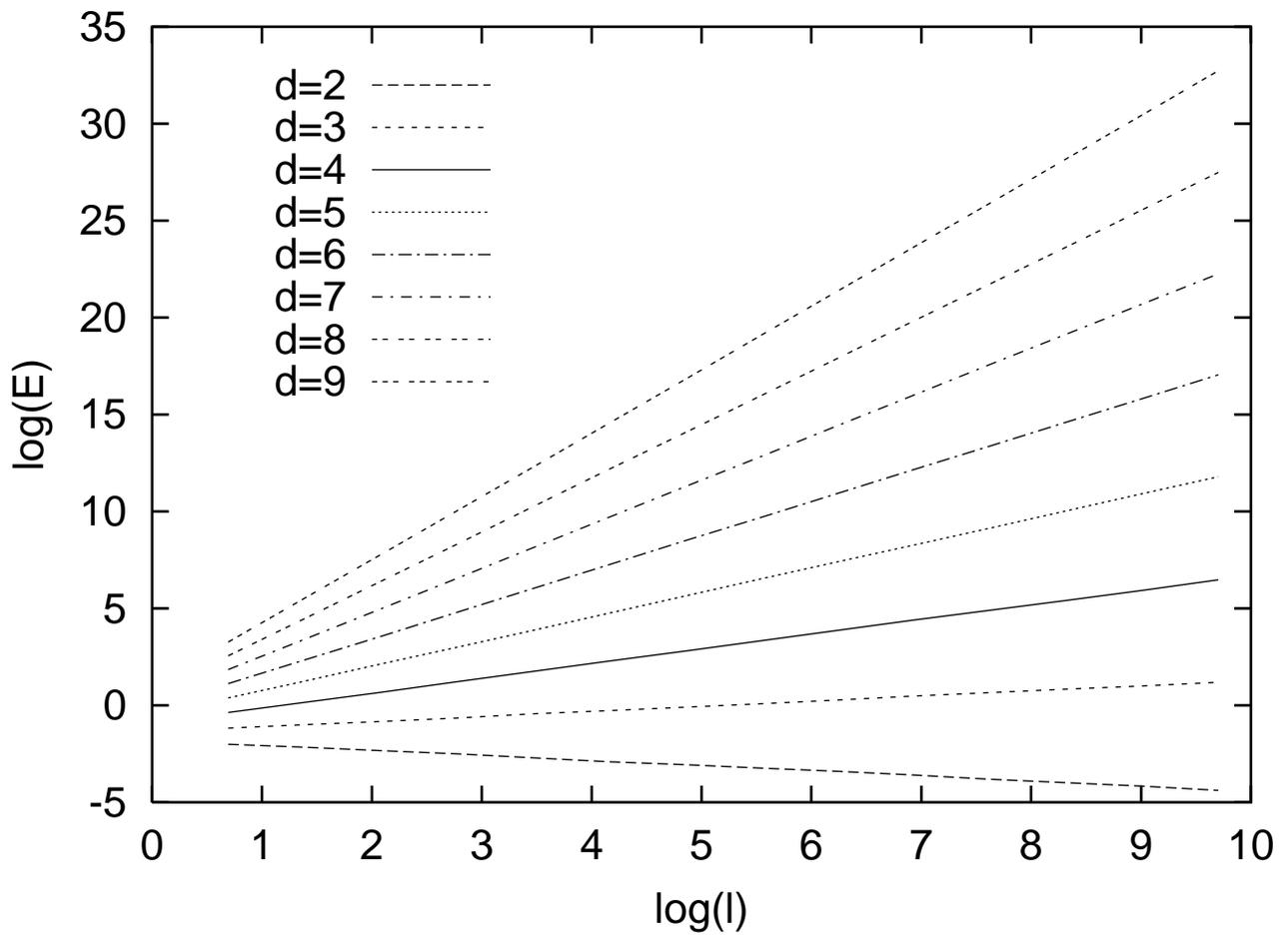
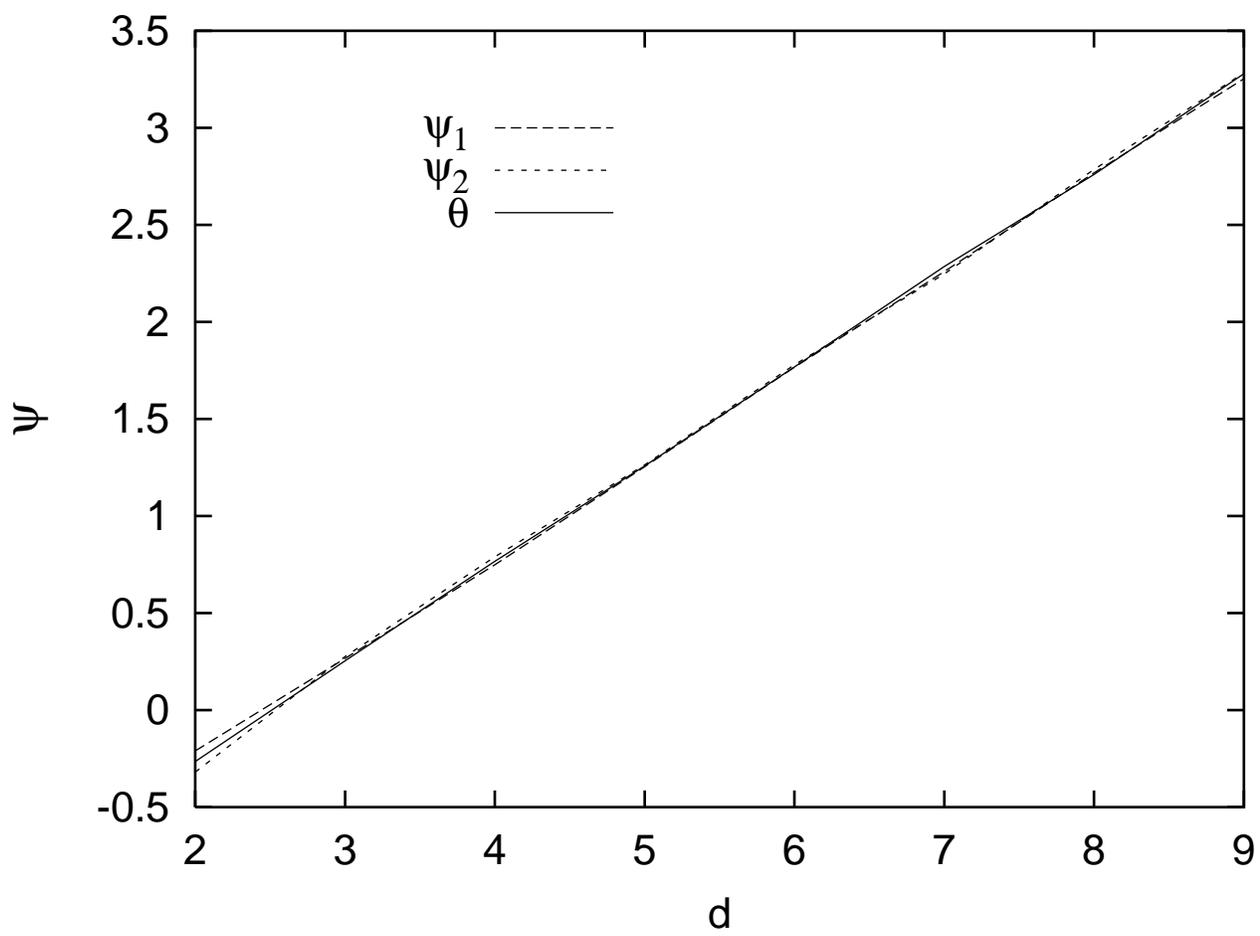
\end{document}